\documentclass[aps,prb,twocolumn]{revtex4-1}
\usepackage{bm}
\usepackage{amsmath,amssymb}
\usepackage{here}
\usepackage[]{graphicx,color}
\usepackage{grffile}

\def\ndmag{Nd$_2$Fe$_{14}$B}

\def\nplcfcnb{(Nd$_{1-\alpha-\beta-\gamma}$Pr$_{\alpha}$La$_{\beta}$Ce$_{\gamma}$)$_{2}$(Fe$_{1-\delta-\zeta}$Co$_{\delta}$Ni$_{\zeta}$)$_{14}$B} 
\def\TC{T_{\mathrm{C}}}

\begin{document}

\title{Data Assimilation Method for Experimental and First-Principles Data:\\
  Finite-Temperature Magnetization of 
  (Nd,Pr,La,Ce)$_{2}$(Fe,Co,Ni)$_{14}$B}

\author{Yosuke Harashima,$^{1}$
  Keiichi Tamai,$^{2}$
  Shotaro Doi,$^{1}$
  Munehisa Matsumoto,$^{2}$
  Hisazumi Akai,$^{2}$\\
  Naoki Kawashima,$^{2}$
  Masaaki Ito,$^{3}$
  Noritsugu Sakuma,$^{3}$
  Akira Kato,$^{3}$
  Tetsuya Shoji,$^{3}$
  and 
  Takashi Miyake$^{1}$
}

\affiliation{
  $^{1}$Research Center for Computational Design of Advanced Functional Materials, 
  National Institute of Advanced Industrial Science and Technology, 
  Tsukuba, Ibaraki 305-8568, Japan
  \\
  $^{2}$The Institute for Solid State Physics, 
  The University of Tokyo, 
  Kashiwa, Chiba 277-8581, Japan
  \\
  $^{3}$Higashifuji Technical Center, 
  TOYOTA MOTOR CORPORATION,
  Susono, Shizuoka 410-1193, Japan
}

\date{\today}

\begin{abstract}
We propose a data-assimilation method for evaluating the finite-temperature magnetization of a permanent magnet 
over a high-dimensional composition space.  
Based on a general framework for constructing a predictor from two data sets including missing values, 
a practical scheme for magnetic materials is formulated 
in which a small number of experimental data in limited composition space are integrated 
with a larger number of first-principles calculation data. 
We apply the scheme to 
\nplcfcnb.
The magnetization in the whole $(\alpha, \beta, \gamma, \delta, \zeta)$ space at arbitrary temperature is obtained. 
It is shown that the Co doping does not enhance the magnetization at low temperatures, 
whereas the magnetization increases with increasing $\delta$ above 320 K.
\end{abstract}

\maketitle

\section{Introduction}
\label{sec:introduction}

Even more than thirty years after the development of neodymium magnet,~\cite{SaFuToYaMa1984,He1991} 
there has still been continuing effort in developing rare-earth permanent magnets. 
One of the central incentives for the development is the resource criticality issue. 
Rare-earth permanent magnets of current industrial use heavily rely on certain rare-earth resources. 
The main phase of the neodymium magnet is {\ndmag} which contains substantial amount of neodymium. 
More critical dysprosium is added to enhance the coercivity, 
while other rare earths such as lanthanum and cerium are abundant. 
Hence more balanced utilization of them is an important issue, 
especially given geologic scarcity and political volatility of the former. 
While Nd$_2$Fe$_{14}$B is famous for its strong saturation magnetization at room temperature, 
it is also known to have relatively low Curie temperature compared to more traditional permanent rare-earth magnets such as Sm$_2$Co$_{17}$. 
Finding a rare-earth magnet with better balance between saturation magnetization and heat resistance thus deserves extensive efforts.~\cite{MiAk2018} 

One of the most common approaches toward improvement of rare-earth permanent magnets has been partial substitution, 
where practitioners replace some elements of a particular mother compound (e.g. Sm$_2$Co$_{17}$ and \ndmag) with other ones.~\cite{Kools2002} 
Even though the ``full'' substitution (e.g. from Nd$_{2}$Fe$_{14}$B to Ce$_2$Fe$_{14}$B) does not work quite well due to the inferior magnetic properties of the resulting compound,~\cite{Pathak2015} 
there still is possibility that better performance compared to the mother compound may be achieved by properly setting the ratio of the substitution. 
This is evidenced by the celebrated Slater-Pauling curve in $3d$ transition metal alloys. 
An example in rare-earth magnets is a recent report for Sm(Fe$_{1-x}$Co$_{x}$)$_{12}$, 
where the magnetization at and above room temperature increases with increasing cobalt concentration.~\cite{HiTaHiHo2017}
These results indicate the importance of searching the optimal chemical composition in a widespread landscape.

The brute-force strategy, however, quickly runs into difficulty as the number of candidates for substituents becomes large. 
This is a typical manifestation of the notorious ``curse of dimensionality'';~\cite{bellman1961curse} 
the number of samples needed for uniform sampling scales exponentially with respect to the number of the candidates. 
Meanwhile, experimental preparation of a permanent magnet involves many time-consuming processes (such as hydrogen decrepitation and annealing), each of which takes one to tens of hours. 
Thus, it is obviously infeasible to experimentally perform uniform and dense study over the entire parameter space in question. 
In order to deal with the curse, practitioners usually restrict their investigations on a rather tiny subset of the parameter space, 
typically by restricting the number of constituent elements~\cite{Alam2013} or by fixing the ratio between elements.~\cite{Tenaud2004} 
While such a treatment is quite useful for studying certain aspects of magnetic compounds, 
it comes at a cost of sampling bias and thereby at a risk of overlooking a truly optimal compound even within the predetermined parameter space. 
Hence, a less biased but still manageable (in terms of the number of experimental trials) approach is highly desirable.

On the other hand, recent development of high-throughput calculation techniques enables us to perform first-principles calculation of thousands of rare-earth magnet compounds.~\cite{korner2016theoretical,nieves2019database,fukazawa2019bayesian}
This is a powerful tool to capture trends over wide parameter space. 
However, it contains a systematic error. 
One remedy we consider in the present work is to perform so-called multitask learning on the data obtained from experiments and those from the first-principles calculations. 
Multitask learning is an approach to improve prediction capability by learning multiple tasks simultaneously (not separately as classical machine learning frameworks do). 
By doing so, one can utilize the hidden relationships among the tasks at hand. 
Given that the computational results are expected to be strongly correlated with experimental ones, 
it is natural to argue that the multitask learning may also work in this case, 
although the first-principles calculations do involve simplifying approximations~\cite{Papanikolaou2002} and hence validity of them should be examined with care. 

In the present work, we overcome the challenge by assimilating a limited number of experimental magnetization data and systematic first-principles calculation. 
The experimental small data is supplemented by the first-principles calculation data, whereas systematic error contained in the latter is corrected by the former. 
While we focus on Nd$_2$Fe$_{14}$B-type compounds (or more specifically, \nplcfcnb) because of their practical relevance, 
we expect that the approach can be easily applied to other types of compounds. 
Apart from the methodology, we present analysis of the saturation magnetization at various temperatures in the entire ($\alpha, \beta, \gamma, \delta, \zeta$) space. 
We show that the magnetization of partially substituted systems considerably deviates from the value linearly interpolated from end points. 
In particular, increase in cobalt concentration enhances the magnetization above 320 K.

The rest of the present paper is organized as follows: 
In Section II, we formulate the present framework of the data assimilation. 
This section also includes the test of the framework using some toy data. 
In Section III, we apply the present framework to magnet compound and discuss the implication of the results. 
We conclude the paper in Section IV.

\section{Data Assimilation: Formalism}
\label{sec:method}

We begin this section by introducing some notations and clarifying the objectives. 
Suppose we would like to model the target variable $\bm{y}\in\mathbb{R}^q$ as a function of the descriptor $\bm{x}\in \mathbb{R}^p$ with some noise $\bm{\varepsilon}$:
\begin{equation}
	\bm{y}=\bm{f}(\bm{x})+\bm{\varepsilon}.
\end{equation}
For example, in the cases discussed in the following sections, 
an element of $\bm{x}$ is a monomial upto the second order power in the concentrations of component elements 
whereas elements of $\bm{y}$ represent the computational and the experimental values of either the magnetization or the Curie temperature.
In the present context, $p$ may or may not be larger than one, but $q$ must, because $\bm{y}$ is supposed to contain both experimental and computational results. 
As for the model $\bm{f}$, we consider a problem of linear regression, on which the multitask learning has been extensively studied:~\cite{Breiman1997}
\begin{equation}
	\bm{f}(\bm{x})=W\bm{x}=\left(\begin{array}{c}\bm{w}_1\cdot\bm{x}\\ \vdots\\ \bm{w}_q\cdot\bm{x} \end{array}\right).
\end{equation}
Now the problem is to estimate the coefficient matrix $W\in\mathbb{R}^{q\times p}$ from given $q$ sets of empirical data $\{\{(\bm{x}_{n;i},y_{n;i} )\}_{n=1}^{N_i}\}_{i=1}^q$. 
In general, values of sampled data $(\bm{x}_{n;i}, y_{n;i})$ and the number of samples $N_{i}$ are not identical among $q$ components. 
One can construct an ordinary least square (OLS) estimator for each $\bm{w}_i (i=1,\cdots,q)$, provided that it exists:
\begin{equation}
	\label{OLSestimates}
	\hat{\bm{w}}_{\mathrm{OLS};i}=(X_i^T X_i)^{-1} X_i^T \bm{Y}_i,\quad 	
\end{equation}
where
\begin{equation}
	X_i=\left(\begin{array}{c}\bm{x}_{1;i}^T\\ \vdots\\ \bm{x}^T_{N_i;i}\end{array}\right), \bm{Y}_i=\left(\begin{array}{c}y_{1;i}\\ \vdots\\ y_{N_i;i}\end{array}\right).
\end{equation}
It is also important to note that the OLS estimator is known to be the best linear unbiased estimator: 
That is, it has the smallest variance among all linear unbiased estimates.~\cite{Hastie2009} 
However,  restricting our interest to unbiased estimates is not necessarily the wisest decision we can make. 
In the present case, the target variables are one physical quantity obtained by different ways: 
experimental and theoretical approaches. 
Ideally, $\bm{w}_{i}$ are equivalent among these ways, but they are not in practice. 
This discrepancy can be resolved by assimilating those data. 
We could construct a slightly biased estimator, with an aid of correlation between multiple outputs $y_i$ $(i=1,\cdots,q)$, to achieve a better tradeoff between bias and variance. 
The focus of the rest of this Section is on the construction of such an estimator.

In order to formulate our approach, 
we hereafter assume that the descriptor $\boldsymbol{x}$ and the target $\boldsymbol{y}$ jointly follow the multivariate Gaussian distribution:
\begin{equation}
\label{eq:mulvalpostulation}
p(\boldsymbol{y},\boldsymbol{x};\Sigma) = \frac{1}{\sqrt{(2\pi)^d|\Sigma|}}\exp\left(-\frac{1}{2}\boldsymbol{z}^T\Sigma^{-1} \boldsymbol{z}\right),
\end{equation}
where $\Sigma$ is a positive-definite (and thereby symmetric) covariance matrix and $\boldsymbol{z}^T:=(\boldsymbol{y}^T,\boldsymbol{x}^T)$.
For later convenience, we use a precision matrix $\Lambda := \Sigma^{-1}$.
The component of $\Lambda$ is represented as 
\begin{equation}
\Lambda = \left(
\begin{array}{cc}
\Lambda_{yy} & \Lambda_{yx}\\
\Lambda_{yx}^T & \Lambda_{xx}\\
\end{array}
\right).
\end{equation}
Then the probability distribution of $\boldsymbol{y}$ conditional on $\boldsymbol{x}$ can be easily found:
\begin{equation}
p(\boldsymbol{y}|\boldsymbol{x};\Lambda)=\sqrt{\frac{|\Lambda_{yy}|}{(2\pi)^d}}\exp\left(-\frac{1}{2}(\boldsymbol{y}-\boldsymbol{\mu})^T\Lambda_{yy} (\boldsymbol{y}-\boldsymbol{\mu})\right),
\end{equation}
where 
\begin{equation}
\label{def:regress-mat}
\boldsymbol{\mu}=-(\Lambda_{yy})^{-1}\Lambda_{yx}\boldsymbol{x}.
\end{equation}

Since Eq.~\eqref{def:regress-mat} defines the regression coefficient matrix of $\boldsymbol{y}$ with respect to $\boldsymbol{x}$, 
the central task here is to estimate the matrix from the data. 
In the present work, we employ maximum likelihood estimation: 
That is, we consider a problem of maximization of the following log likelihood function $L(\Lambda|\{(\boldsymbol{x}_n,\boldsymbol{y}_n)\}_{n=1}^N)$: 
\begin{equation}
L(\Lambda_{yy},\Lambda_{yx}|\{(\boldsymbol{x}_n,\boldsymbol{y}_n)\}_{n=1}^N)=\sum_n \log p(\boldsymbol{y}_n|\boldsymbol{x}_n;\Lambda).
\end{equation}
Here we note that, even though the original precision matrix $\Lambda$ has $(p+q)(p+q+1)/2$ independent elements, 
optimization of only $\Lambda_{yy}$ and $\Lambda_{yx}$ suffices as far as estimation of $\boldsymbol{\mu}$ $\Lambda_{yy}$ is concerned. 

Until here, the present formulation is fairly general (under the assumption of (\ref{eq:mulvalpostulation}), at least), 
and we did not assume any further relations between the target variables. 
Although combining multiple measurements may be beneficial for a better estimation (in terms of combined error) even when underlying mechanisms of those outputs are completely independent,~\cite{Stein1956,Efron1975} 
it is usually advisable to incorporate a relation between the outputs during the estimation process (when reasonably possible). 
This is particularly true when data for some target variables are substantially harder to gather than the others but one has a good guess over the relation between these two.

In order to introduce a bias (correlation) to our estimator, 
we postulate that the experimental values $y_{\mathrm{expt}}(\boldsymbol{x})$ can be represented by a scalar multiplication of the prediction $y_{\mathrm{comp}}(\boldsymbol{x})$ derived from computational data 
and a residual part $R(\boldsymbol{x})$ which involves substantially fewer terms than those originally considered for modeling the computational results 
(hereafter we assume $q=2$, and refer to $y_1$ as computational output $y_{\mathrm{comp}}$ and $y_2$ as experimental output $y_{\mathrm{expt}}$).
\begin{align}
	\label{expdecomposition}
	y_{\mathrm{expt}}(\boldsymbol{x}) &= Cy_{\mathrm{comp}}(\boldsymbol{x})+R(\boldsymbol{x}), 
\end{align}
where
\begin{align}
        R(\boldsymbol{x}) &:=\boldsymbol{w}_{\mathrm{res}}\cdot\boldsymbol{x}_{\mathrm{proj}}.
\end{align}
Here, $\boldsymbol{w}_{\mathrm{res}}\in \mathbb{R}^r$ $(r<p)$ and $\boldsymbol{x}_{\mathrm{proj}}$ is a natural projection of $\boldsymbol{x}$ onto the parameter space of relevant descriptors. 
This assumption can be easily implemented on the present formulation by enforcing 
\begin{equation}
\Lambda_{y_{\mathrm{expt}},i} = 0\quad \mathrm{for}\quad i\in \mathrm{irrelevant\ descriptors}.
\end{equation}
This indicates that $y_{\mathrm{expt}}$ only indirectly correlates with the $i$-th descriptor through other variables.
In this case, $C$ in Eq.\ (\ref{expdecomposition}) can be expressed in terms of $\Lambda_{yy}$:
\begin{equation}
	C=\Lambda_{y_{\mathrm{expt}},y_{\mathrm{comp}}}/\Lambda_{y_{\mathrm{comp}},y_{\mathrm{comp}}}.
\end{equation}

While the description of the present framework is conceptually complete up to here, we have to face with one more complication in practice. 
A central concern here is that we usually have the smaller number $N_{\mathrm{expt}}$ of experimental data than that $N_{\mathrm{comp}}$ of computational ones, 
thus, some data have a missing value in either $y_{\mathrm{expt}}$ or $y_{\mathrm{comp}}$. 
Although one may simply discard such missing pairs during the analysis, 
one could not benefit from abundance of computational data in this approach and hence the estimates could not be very efficient.

In order to address this issue, we use the direct likelihood. 
The idea behind the direct likelihood is that we integrate out the missing variable. 
The direct likelihood is written as 
\begin{align}
  &L(\Lambda_{yy},\Lambda_{yx}|\{(\boldsymbol{x}_n,\boldsymbol{y}_n)\}_{n=1}^N)= \displaystyle \sum_{n\in \Omega_{\mathrm{comp},\mathrm{expt}}} \log p(\boldsymbol{y}_n|\boldsymbol{x}_n;\Lambda) \nonumber \\
  & \hspace{60pt} \displaystyle + \sum_{n\in \Omega_{\mathrm{comp}}} \log p_{\mathrm{comp}}(y_{\mathrm{comp},n}|\boldsymbol{x}_n;\Lambda) \nonumber \\
  & \hspace{60pt} \displaystyle + \sum_{n\in \Omega_{\mathrm{expt}}} \log p_{\mathrm{expt}}(y_{\mathrm{expt},n}|\boldsymbol{x}_n;\Lambda)
  \label{eq:likelihood}
\end{align}
by decomposing a sample set into three subsets $\Omega_{\mathrm{comp},\mathrm{expt}}$ where both the experimental and computational data are available, 
and $\Omega_{\mathrm{expt}}$, $\Omega_{\mathrm{comp}}$ where either of the two is missing 
(for example, samples in $\Omega_{\mathrm{expt}}$ only contain experimental data).
The distributions for the missing data are
\begin{align}
& p_{\Gamma}(y_{\Gamma}|\boldsymbol{x};\Lambda)=\int \mathrm{d}y_{\bar{\Gamma}}\;p(y_{\Gamma},y_{\bar{\Gamma}}|\boldsymbol{x};\Lambda)
\\
& \quad = \sqrt{\dfrac{\left|\bar{\Lambda}_{\Gamma\Gamma}\right|}{2\pi}}
\exp\left(-\dfrac{1}{2}\left(y_{\Gamma}-\mu_{\Gamma}\right)\bar{\Lambda}_{\Gamma\Gamma}\left(y_{\Gamma}-\mu_{\Gamma}\right)\right),
\end{align}
where
\begin{equation}
\bar{\Lambda}_{\Gamma\Gamma} := \Lambda_{\Gamma\Gamma}-\Lambda_{\Gamma\bar{\Gamma}}\Lambda_{\bar{\Gamma}\bar{\Gamma}}^{-1}\Lambda_{\bar{\Gamma}\Gamma},
\end{equation}
and a pair of $(\Gamma,\bar{\Gamma})$ represents either $(\mathrm{expt},\mathrm{comp})$ or $(\mathrm{comp},\mathrm{expt})$.
Use of the direct likelihood is justified by the fact that the missingness of data is simply a matter of design of the measurements in this case and hence unrelated to the values that have been missed 
(in other words, data are missing at random (albeit not completely) in the sense of Rubin~\cite{Rubin1976}).

Then we optimize $\Lambda_{yy}$ and the relevant part of $\Lambda_{yx}$ for Eq.~\eqref{eq:likelihood}.
The optimization problem can be computationally solved using the limited-memory Broyden-Fletcher-Goldfarb-Shanno algorithm \cite{Broyden1970,Fletcher1970} with a simple box constraints (L-BFGS-B).~\cite{Byrd1995}

\begin{figure}[ht]
\includegraphics[width=\hsize]{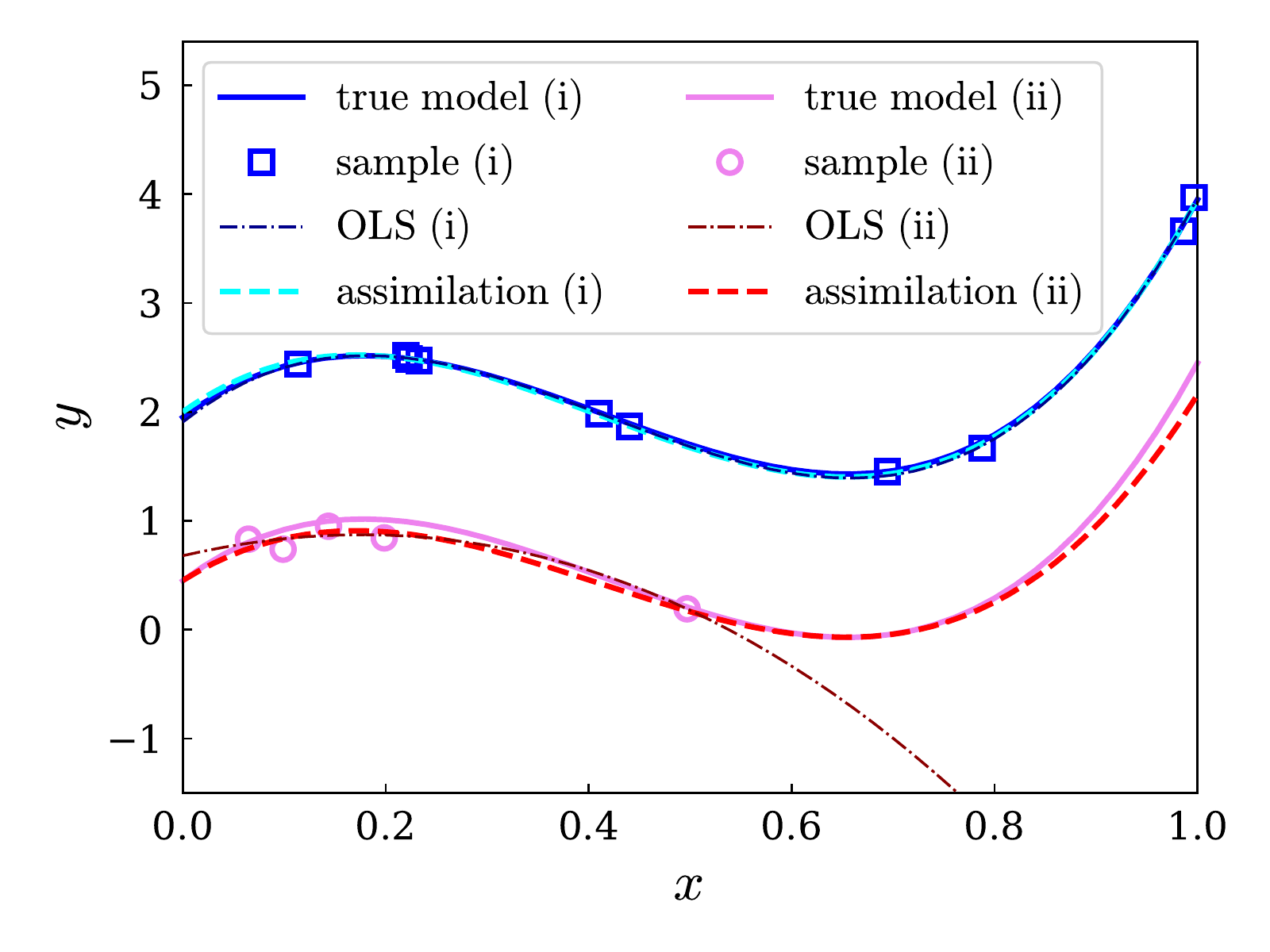}
\caption{(Color online) An example for the data assimilation. 
The true models are given by hand as (i) (blue solid curve) and (ii) (violet solid curve). 
Sample data generated from the true models are shown as blue squares and violet points. 
The prediction from the data assimilation is shown as red for the model (i) and cyan for the model (ii). 
The prediction from the OLS fit is also shown as a dot-dashed line for comparison.}
\label{fig:toydata}
\end{figure}

To see features of the method, we applied it to a toy data. 
The toy data was generated from two true models,
\begin{equation}
	f_\mathrm{i}(x)=2-x+5(x-0.7)^2+20(x-0.5)^3,
\end{equation}
\begin{equation}
	f_\mathrm{ii}(x)=f_\mathrm{i}(x)-1.5,
\end{equation}
where $x$ is an input parameter and the descriptor consists of $(1,x,x^2,x^3)$. 
The data (i) and (ii) mimic computational and experimental data, respectively.
10 samples for model (i) and 5 for model (ii) were sampled randomly. 
Noise was applied with widths 0.03 for $y_{\mathrm{i}}$ and 0.1 for $y_{\mathrm{ii}}$, respectively. 
An example sample data and a result of the assimilation are shown in Fig.~\ref{fig:toydata}. 
Although the sampled values of descriptors do not match between the two models, the prediction well agree with the true model. 
We also compared with a simple least square fitting for these models separately. 
For the model (i), the OLS fit agrees with the true model as well as the assimilation. 
For the model (ii), however, the OLS fit disagrees with the true model for $0.5 < x < 1.0$, whereas, the assimilation well predicts the true model even for that region.

The remaining issue is to determine the list of candidates for descriptors. 
Since, as we will see in Section~\ref{sec:application}, 
the behavior of the quantities of interest (e.g.\ magnetization, Curie temperature) is rather simple and we do not expect to encounter a singular point where these quantities diverge, 
it suffices to model them by polynomial functions of the concentration of each element. 
More specifically, we modeled the first-principles results by a second-order polynomial function:
\begin{equation}
y_{\mathrm{comp}}=\sum_{i_{\alpha}+i_{\beta}+i_{\gamma}+i_{\delta}+i_{\zeta}\le 2}c_{i_{\alpha}i_{\beta}i_{\gamma}i_{\delta}i_{\zeta}}\alpha^{i_{\alpha}}\beta^{i_{\beta}}\gamma^{i_{\gamma}}\delta^{i_{\delta}}\zeta^{i_{\zeta}}, 
\label{eq:y_comp}
\end{equation} 
where $\alpha$, $\beta$, $\gamma$, $\delta$, and $\zeta$ denote respectively the concentration of Pr, La, Ce, Co, and Ni.
As for the list of candidates for relevant descriptors for $R$ in Eq.~\eqref{expdecomposition}, 
we considered a constant term for the magnetization, and constant and $\delta$ linear terms for the Curie temperature.

\section{Magnetization of (N\lowercase{d},P\lowercase{r},L\lowercase{a},C\lowercase{e})$_2$(F\lowercase{e},C\lowercase{o},N\lowercase{i})$_{14}$B}
\label{sec:application}

\subsection{Data assimilation for magnetization}
\label{subsec:MT}
Now that we have introduced a general procedure for the data assimilation, 
the next step is to apply the methodology to build up a flexible framework for predictions of the essential properties of the magnetic compounds. 
By `flexibility,' we mean the ability to predict the properties at an arbitrary temperature, in addition to one for arbitrary combination of the doping concentrations. 

A problem here is that both the experiments and first-principles calculations suffer from their limitations. 
On one hand, experiments are hard to perform at an arbitrary condition due to e.g. difficulty in synthesis and limitation in experimental facilities. 
On the other hand, density functional theory (DFT),~\cite{HoKo1964,KoSh1965} on which our first-principles calculation are based, rely on an approximation to the exchange-correlation functional in practical applications. 
Furthermore, DFT only address the ground state property of the system (in other words, the system at 0 K). 
Finite-temperature magnetism such as Curie temperature can be evaluated 
by combining DFT with a mean field theory of classical spin dynamics.\cite{Rosengaard1997,Pajda2001} 
However, quantitative agreement with experiments is limited. 
Magnetism is a consequence of quantum many-body effect, which requires sophisticated theoretical treatment. 
Therefore, the experiments and the calculations are in a sense complementary to each other.

\begin{figure}[ht]
  \includegraphics[width=\hsize]{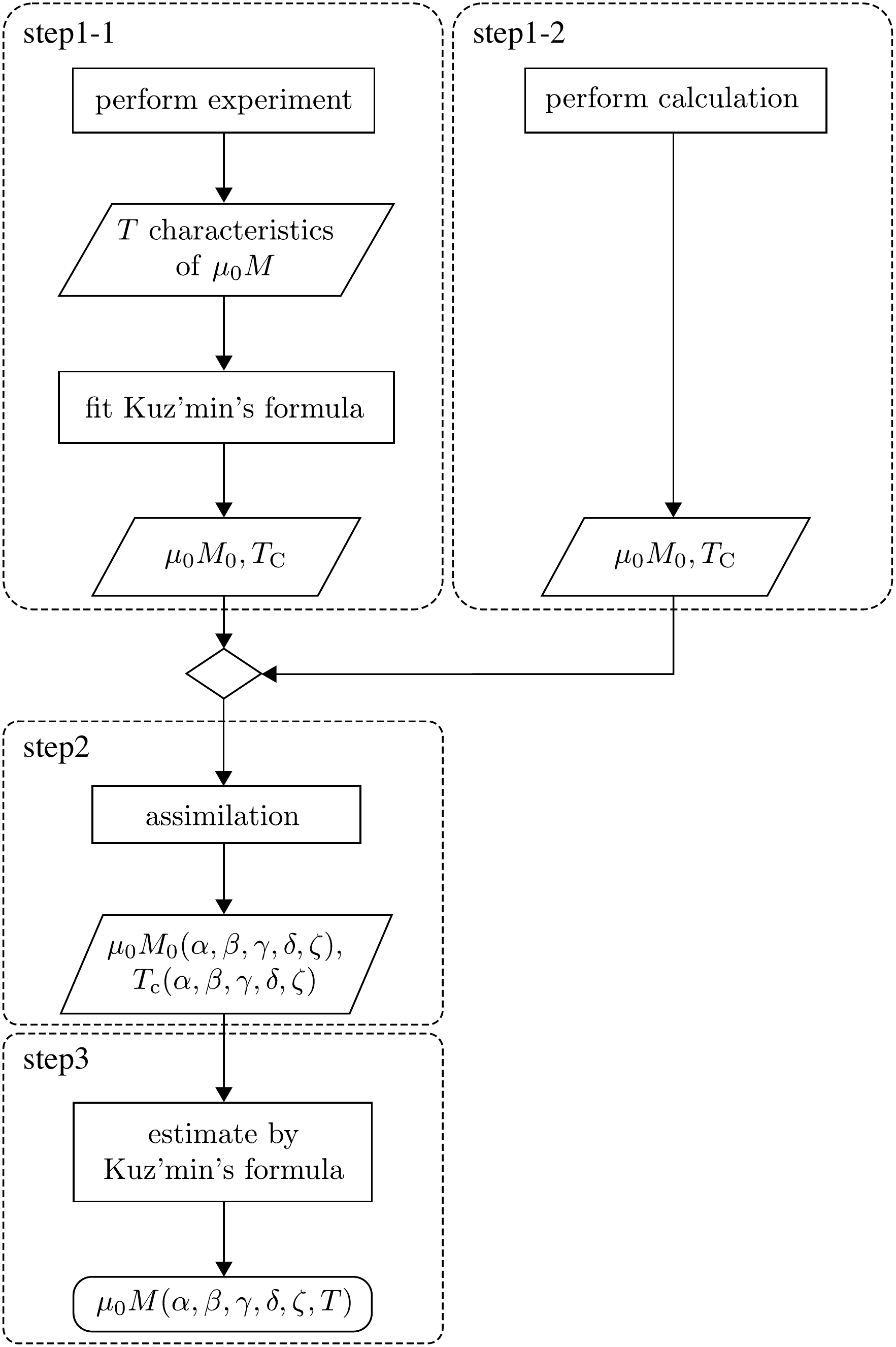}
  \caption{Flowchart of data-assimilation method for finite-temperature magnetization.}
  \label{fig:flowchart}
\end{figure}

\begin{figure}[ht]
  \includegraphics[width=\hsize]{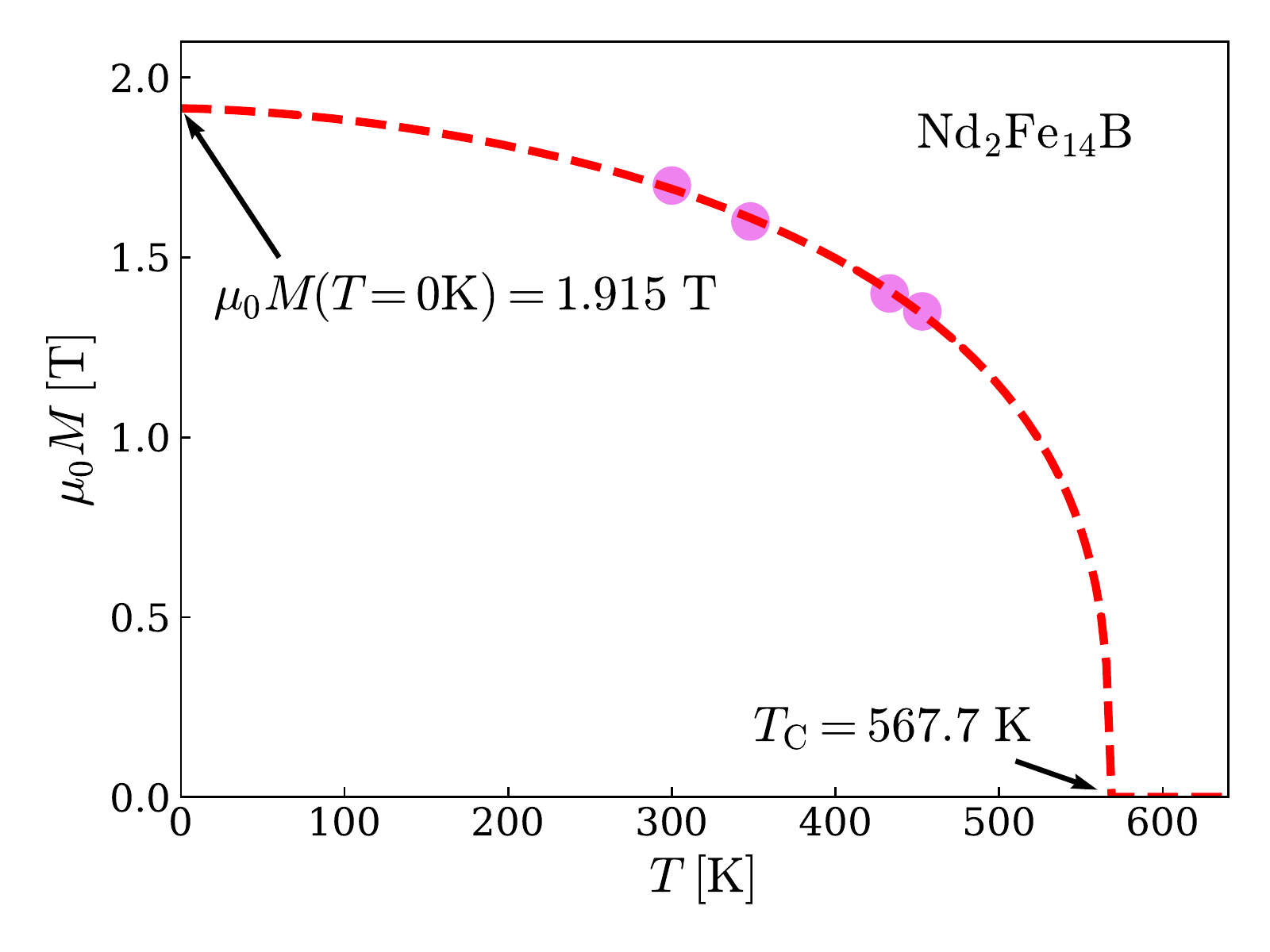}
  \caption{(Color online) The magnetization of Nd$_{2}$Fe$_{14}$B as a function of temperature. 
    The violet points denote the experimental observation.
    The red dashed curve is a regression curve derived from a least square fitting to Eq.~\eqref{eq:Kuzmin}.
    The obtained values for $\mu_{0}M(T=0\mathrm{K})$ and $\TC$ are also shown. 
  }
  \label{fig:kuzmin_fit}
\end{figure}

To fix this problem, we introduce a practical scheme to estimate the finite-temperature magnetization.  
The scheme consists of the following four steps and its flowchart is illustrated in Fig.~\ref{fig:flowchart}.

{\bf Step 1-1:}
We prepared 119 experimental samples of {\nplcfcnb }, each of which has different chemical composition 
(this includes one sample value extracted from Refs.~\onlinecite{Hock1988,Grossinger1985}). 
The magnetization $\mu_0 M(T)$ of each sample was measured at 3--7 temperatures within a range from 300 K to 473 K. 
Experimental details can be found in Appendix~\ref{sec:expdetails}. 
We converted $\mu_0 M(T)$ into the Curie temperature $\TC$ and the magnetization $\mu_0M(T=0)$ at 0 K, 
which can then be directly compared to the computational results. 
To do this, we referred to Kuz'min's empirical formula for magnetization:~\cite{Ku2005}
\begin{equation}
	\label{eq:Kuzmin}
	\mu_0 M(T)=\mu_0 M_{0} [1-st^{\frac{3}{2}}-(1-s) t^{\frac{5}{2}}]^{\frac{1}{3}},
\end{equation}
where $t:=T/\TC$, $\mu_{0} M_{0}:=\mu_{0} M(T=0)$, and $s$ is a phenomenological shape parameter specific to the compound.~\cite{kuz2005temperature,kuz2006factors} 
For each chemical composition, $\mu_{0}M_{0}$ and $\TC$ were obtained by the least square fitting. 
Following Ref.~\onlinecite{ItYaDeGi2016}, we fixed $s$ to be 0.6 in the present work. 
(To be more accurate, $s$ could be fitted as well as $\mu_{0}M_{0}$ and $\TC$.)

{\bf Step 1-2:}
On the theoretical side, we calculated the $\mu_0 M_{0}$ and $\TC$ following the method explained in Appendix~\ref{sec:calcdetails}. 
The calculations were performed for 2869 compositions uniformly distributed in the $(\alpha, \beta, \gamma, \delta, \zeta)$ space. 

{\bf Step 2:}
We then applied the data-assimilation method to the obtained $\mu_{0}M_{0}$ and $\TC$. 
We adopted the following regression models: 
\begin{align}
\mu_0 M_{0}^{\rm comp}(\alpha, \beta, \gamma, \delta, \zeta) &= \sum_{I} M_{I}^{\rm comp} \alpha^{i_\alpha} \beta^{i_\beta} \gamma^{i_\gamma} \delta^{i_\delta} \zeta^{i_\zeta}, \\
\TC^{\rm comp}(\alpha, \beta, \gamma, \delta, \zeta) &= \sum_{I} T_{I}^{\rm comp} \alpha^{i_\alpha} \beta^{i_\beta} \gamma^{i_\gamma} \delta^{i_\delta} \zeta^{i_\zeta}, 
\end{align}
where $I := (i_\alpha,i_\beta,i_\gamma,i_\delta,i_\zeta)$. 
Both quantities were expanded up to the quadratic terms, namely $0 \leq \sum_j i_j \leq 2$.
We assumed that the experimental data are correlated with the computational data as follows: 
\begin{align}
\mu_0 M_{0}^{\rm expt}(\alpha, \beta, \gamma, \delta, \zeta) &= C_{m} \mu_0 M_{0}^{\rm comp}(\alpha, \beta, \gamma, \delta, \zeta) + m_0, 
\\
\TC^{\rm expt}(\alpha, \beta, \gamma, \delta, \zeta) &= C_{t} \TC^{\rm comp}(\alpha, \beta, \gamma, \delta, \zeta) + t_0 + t_1 \delta. 
\label{eq:TCcorrelation}
\end{align}
The coefficients $\{M_{I}^{\rm comp}\}_{I}$, $C_{m}$, $m_0$ and $\{T_{I}^{\rm comp}\}_{I}$, $C_{t}$, $t_0$, $t_1$ were determined by 
optimizing the likelihood Eq.~\eqref{eq:likelihood} in terms of the parameters of the multivariate normal distribution $\Lambda_{yy}$ and $\Lambda_{yx}$.
Then, the coefficients were given by $-(\Lambda_{yy})^{-1}\Lambda_{yx}$ as mentioned at Eq.~\eqref{def:regress-mat}.

In general, we may include higher-order terms to make a prediction model more flexible and accurate, 
which brings complexity in the determination of the coefficients in the prediction model. 
In the present case, we included the linear term of $\delta$ in Eq.~\eqref{eq:TCcorrelation} and we will discuss that point later.

{\bf Step 3:}
The above data assimilation gives model functions for 
$\mu_0 M_0=\mu_0 M_{0}^{\rm expt}$ and $\TC=\TC^{\rm expt}$ at arbitrary composition. 
Applying Kuz'min's formula Eq.~\eqref{eq:Kuzmin} to these data, we estimate $\mu_0 M$ at finite temperature. 
In this way, we can predict the experimental magnetization at arbitrary point in the $5+1$ dimensional space 
spanned by $\alpha, \beta, \gamma, \delta, \zeta$, and $T$. 

\subsection{Results and discussion}
\label{subsec:result}

\begin{figure*}[ht]
  \includegraphics[width=\hsize]{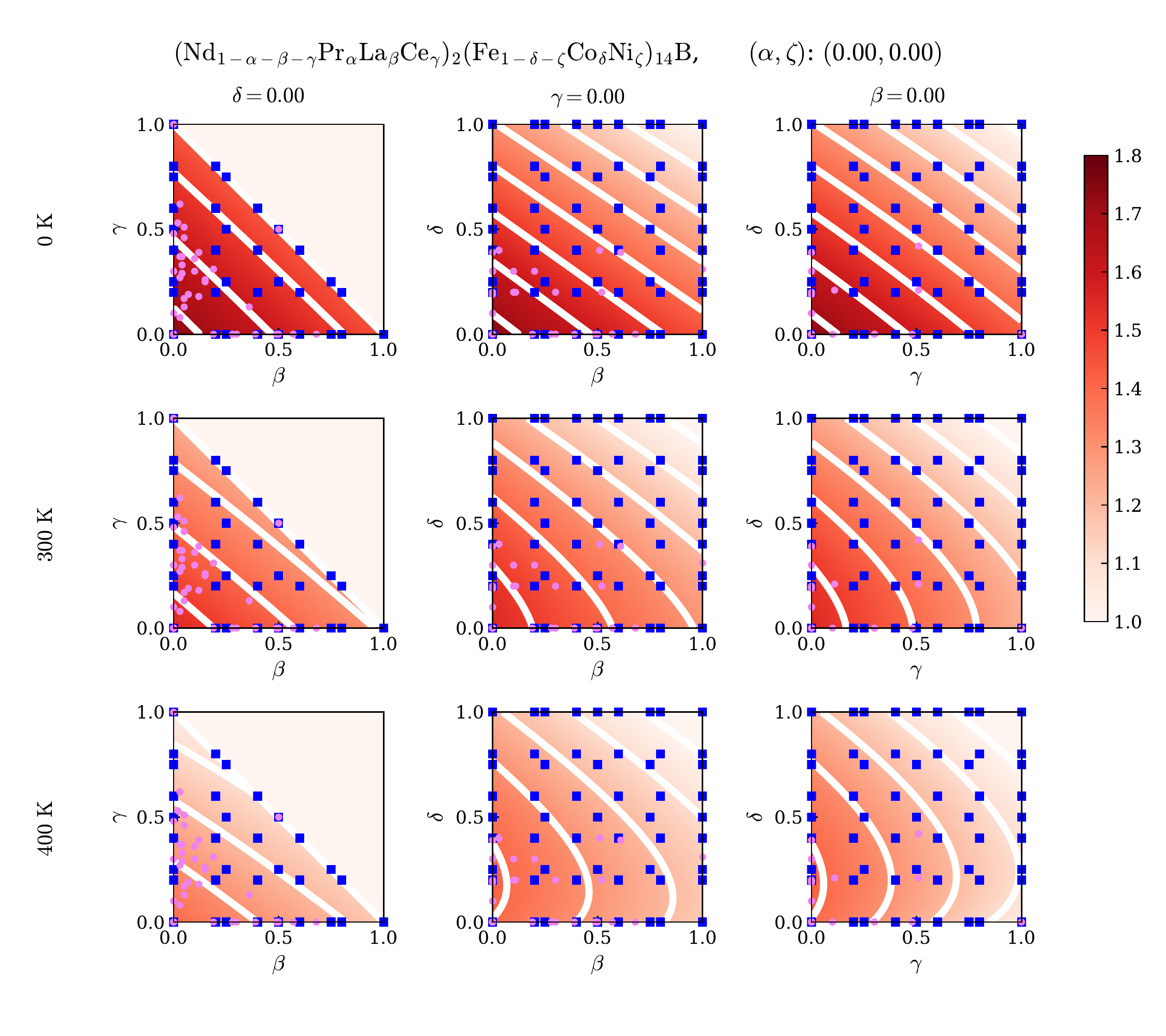}
  \caption{(Color online) Color map of the magnetization $\mu_{0}M(\alpha=0,\beta,\gamma,\delta,\zeta=0,T)$ at $T=$ 0, 300, and 400 K.
    Experimental values, estimated by the data-assimilation method, on $(\beta,\gamma,\delta=0)$, $(\beta,\gamma=0,\delta)$, and $(\beta=0,\gamma,\delta)$ planes are shown.
    The values of the magnetization are described in tesla.
    Blue squares are sampling points of the first-principles calculation data and 
    violet points are those of the experimental data.
    Contour lines are also shown with interval of 0.1 T.
    Note that the upper triangle region on the $(\beta, \gamma)$ plane is irrelevant, 
    because $\beta+\gamma$ cannot exceed 1. 
  }
  \label{fig:magnetization_colormap}
\end{figure*}

\begin{figure*}[ht]
  \includegraphics[width=\hsize]{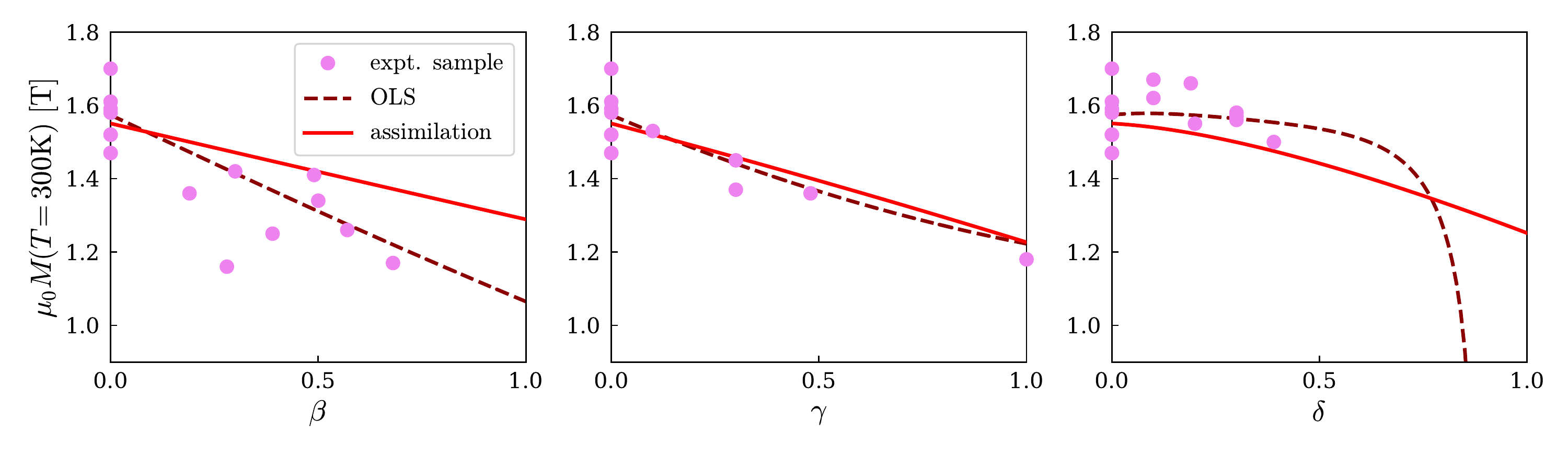}
  \caption{(Color online) Prediction (red solid lines) for the magnetization $\mu_{0}M$ at $T$=300 K is shown 
    along $(0,\beta,0,0,0)$, $(0,0,\gamma,0,0)$, and $(0,0,0,\delta,0)$.
    Corresponding result for the OLS fit is shown as darkred dashed lines.
    Violet points indicate the experimental sample data observed at 300 K.
  }
  \label{fig:magnetization_300K}
\end{figure*}

Figure~\ref{fig:magnetization_colormap} shows color maps of the magnetization on 
$(\beta, \gamma)$, $(\beta, \delta)$, $(\gamma, \delta)$ planes at 0, 300, and 400 K, 
where other variables in $\alpha, \beta, \gamma, \delta, \zeta$ are set to zero in each case. 
We see that the contour lines are not straight. This means that the magnetization varies non-linearly in the composition space.
The experimental data are unevenly distributed, 
which are interpolated and extrapolated over the wide composition space 
with the help of computational data by the data-assimilation method. 

Figure~\ref{fig:magnetization_300K} is a comparison of $\mu_{0}M$ at $T$=300 K 
between measured values and prediction by data assimilation and OLS fit 
along the $\beta$, $\gamma$, and $\delta$ axes.
Estimated root mean square errors between the predictions and the experimental data at 300 K are 0.085 T for the assimilation and 0.067 T for the OLS fit. 
It can be seen that both the predictions are able to predict the sampled experimental data, 
however, the OLS-fit model extremely decreases at the high $\delta$ region where sample points are absent.
This seems to be typical behavior of overfitted models in which sampled data are well described while prediction is sacrificed. 
On the other hand, the extreme decrease is not observed in the present assimilation method. 

\begin{table}[ht]
  \caption{Root mean square error for the assimilated predictions, $\TC$ and $\mu_{0}M_{0}$, 
    corresponding to the components for the experiment. 
    The error for both the 10-fold cross validation (CV) and the extrapolation are shown.
    In the extrapolation, the test data is the experimental data in the range of $\alpha \geq 0.5$.
    For comparison, results of the OLS fit are also shown.
  }
  \vspace{5pt}
  \begin{center}
    \begin{tabular}{lccp{50pt}p{75pt}}
      \hline \hline
      & & & \hfil OLS fit \hfil & \hfil data assimilation \hfil
      \\
      \hline
      10-fold CV    & $\TC$    & [K] & \hfil 172.5  \hfil & \hfil 43.8 \hfil
      \\
                    & $\mu_{0}M_{0}$ & [T] & \hfil 0.719 \hfil & \hfil 0.092 \hfil
      \\
      extrapolation & $\TC$    & [K] & \hfil 854.6 \hfil & \hfil 46.8 \hfil
      \\
                    & $\mu_{0}M_{0}$ & [T] & \hfil 6.180 \hfil & \hfil 0.097 \hfil
      \\
      \hline \hline
    \end{tabular}
  \end{center}
  \label{table:error}
\end{table}

In order to examine the accuracy of the data-assimilation method in comparison with the straightforward OLS fit, 
we performed 10-fold cross-validation for the experimental $\mu_{0}M_{0}$ and $\TC$. 
The experimental data were randomly divided into 10 groups.
One of the groups was taken as a test data and the remainings were used for training.
The data assimilation was applied to the training data along with the whole computational data, 
then the root mean square error between the predicted experimental value and the test data was evaluated.
10 error values were obtained by changing test data to another group.
We iterated 10 sets of this procedure, 
and took an average of the errors. 
For the OLS fit, we used the training data only (without computational data). 
As demonstrated in the upper half of Table~\ref{table:error}, 
the average error of the present methodology was found to be significantly smaller than that of the OLS fit, 
indicating high generalizability of the former. 
The advantage of the data assimilation compared to the OLS fit is more drastic when the test data are distributed 
outside the region of training data. 
To see how the data assimilation works in the case of ``extrapolation'', we extracted the experimental data with $\alpha > 0.5$ as test data 
and using the remaining data as training data.
The data assimilation and the OLS fit were applied to the training data as done for the cross validation.
The estimated error is shown in the lower half of Table~\ref{table:error}. 
It indicates that the data assimilation extrapolates both $\mu_{0}M_{0}$ and $\TC$ within permissible ranges of errors, 
while the OLS fit breaks down. 

Coming back to Fig.~\ref{fig:magnetization_colormap}, 
the predicted magnetization gives us useful information for Co substitution. 
The magnetization monotonically decreases with increasing $\delta$ at $T=0$ K and 300 K, 
whereas a Slater-Pauling like peak is found as a function of $\delta$ at 400 K. 
This is in sharp contrast with bcc-(Fe,Co), where Co doping up to $\sim$30 \% enhances the magnetization 
in the whole temperature region. 
This indicates that Co substitution enhances the magnetization at high temperatures 
via enhancement of the Curie temperature. 
In Sm(Fe,Co)$_{12}$, 
such reduction of the magnetization at zero temperature and enhancement of the Curie temperature by Co doping have been also observed in experiment.~\cite{HiTaHiHo2017}

\begin{figure}[ht]
  \includegraphics[width=\hsize]{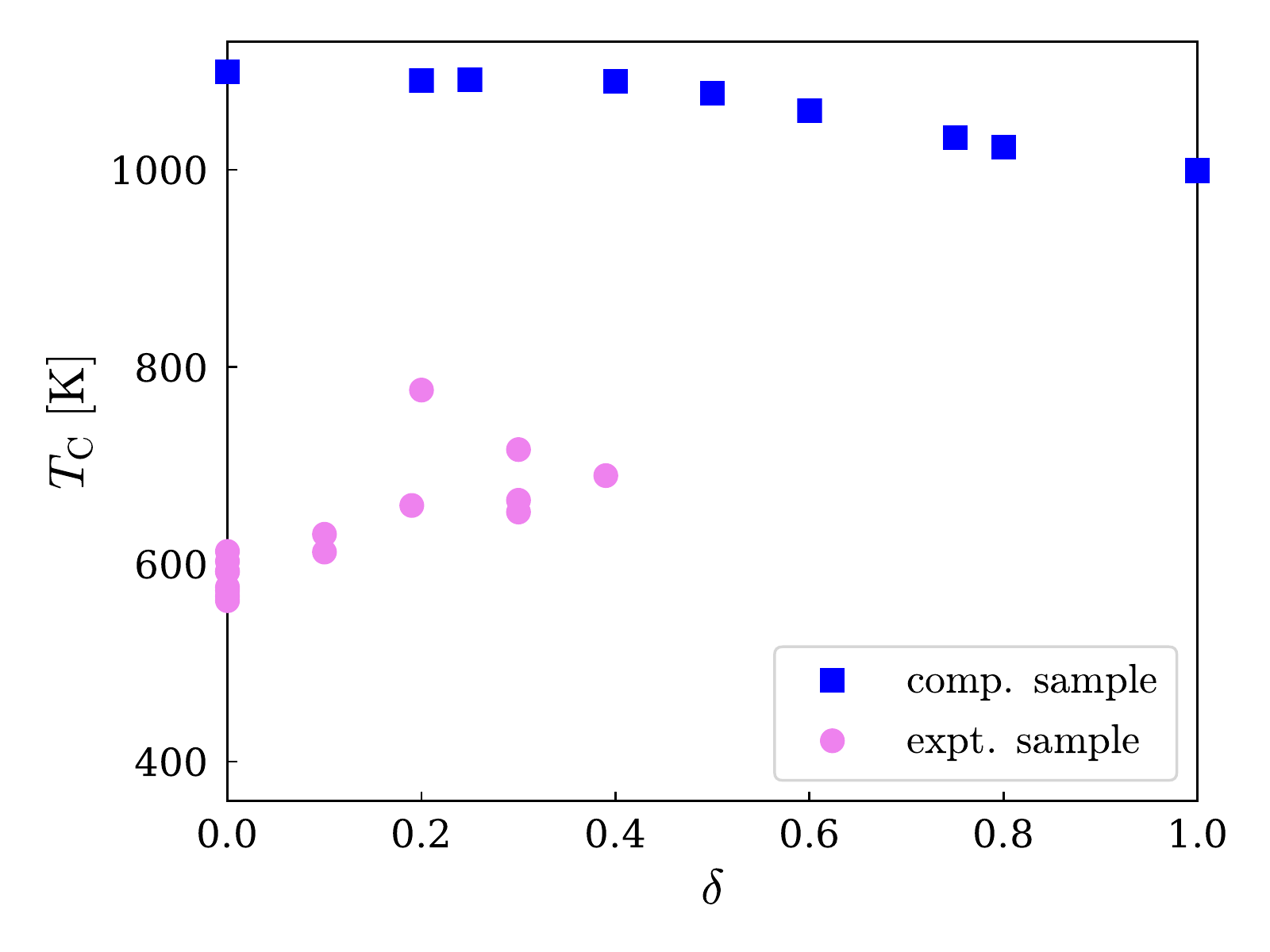}
  \caption{(Color online) Co concentration dependence of the Curie temperature of Nd$_{2}$(Fe$_{1-\delta}$Co$_{\delta}$)$_{14}$B.
    Blue squares denote the first-principles data and 
    violet points denote the experimental data.
  }
  \label{fig:tc_calc_exp}
\end{figure}

\begin{figure*}[ht]
  \includegraphics[width=0.9\hsize]{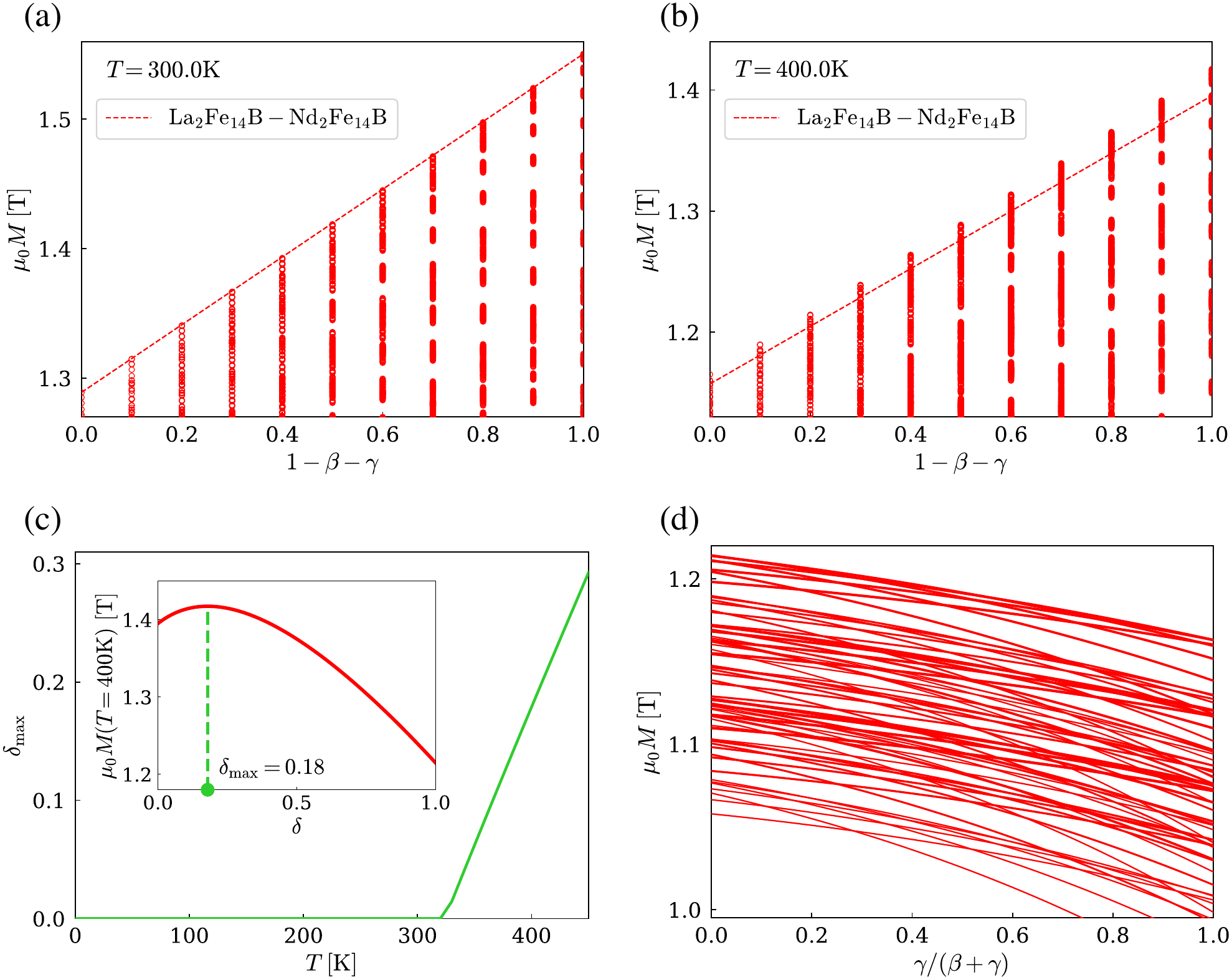}
  \caption{(Color online) 
Scatter plots of the magnetization at (a) 300 K and (b) 400 K as a function of Nd and Pr concentration ($1-\beta-\gamma$).
The dotted straight line connects the magnetizations of two systems: La$_{2}$Fe$_{14}$B and Nd$_{2}$Fe$_{14}$B. 
(c) Cobalt concentration associated with chemical composition giving maximum magnetization. The horizontal axis is temperature in kelvin. 
The inset is an example of the magnetization at 400 K of Nd$_{2}$(Fe$_{1-\delta}$Co$_{\delta}$)$_{14}$B. A green point denotes the $\delta_{\mathrm{max}}$ at the temperature.
(d) Magnetization at 400 K as a function of $\gamma / (\beta + \gamma)$. Other variables are fixed in each line and sampled within the range of  
$0.0 \leq \alpha \leq 0.2$, $0.8 \leq \beta + \gamma \leq 1.0$, $0.0 \leq \delta \leq 0.3$, and $0.0 \leq \zeta \leq 0.1$.
  }
  \label{fig:magnetization_scatter_plot}
\end{figure*}

This conclusion is based on the fact that the $\TC$ increases with the Co concentration. 
However, it contains a tricky problem in the first-principles calculation. 
Figure~\ref{fig:tc_calc_exp} shows the Co concentration dependence of $\TC$ obtained by the experiment (step1-1 in Fig.~\ref{fig:flowchart}) and by first-principles calculation (step1-2). 
As seen in the figure, the calculated value decreases as the concentration increases, 
i.e., the trend is opposite from that of experiment.
This discrepancy originates from theoretical errors in the first-principles data, 
which are systematic errors but not accidental ones. 
It is a hard task to reduce the errors by improving approximations contained in theoretical methods. 
However, such systematic errors can be complemented by the data assimilation 
if there is a correlation between computational data and experimental data, even though the correlation is negative.
In our scheme, the negative correlation in ${\TC}$ is expressed by the linear term in the right side of Eq.~\eqref{eq:TCcorrelation}.

Figure \ref{fig:magnetization_scatter_plot} (a) and (b) show scatter plots of the magnetization at 300 and 400 K, respectively, 
where the predicted magnetizations on uniform mesh points are plotted against the sum of Nd and Pr concentration ($1-\beta-\gamma$), 
which can be regarded as the ratio of critical rare earths in the present system. 
The dotted straight line connects the magnetizations of two systems: La$_{2}$Fe$_{14}$B and Nd$_{2}$Fe$_{14}$B. 
While all the points are below the dotted line at 300 K, some points appear above the line at 400 K. 
This is attributed to the enhancement of the magnetization at high temperatures by Co doping. 
The maximum magnetization at $T=400$ K is achieved in Nd$_{2}$(Fe,Co)$_{14}$B with the cobalt concentration $\delta_{\rm max}=$ 0.18. 
Figure~\ref{fig:magnetization_scatter_plot} (c) shows the temperature dependence of $\delta_{\rm max}$. 
The $\delta_{\rm max}$ is non-zero above 320 K, and increases with raising the temperature. 
Finally, we analyze the magnetization as a function of Ce concentration. 
In Fig. \ref{fig:magnetization_scatter_plot} (d), the magnetization is plotted as a function of $\gamma / (\beta + \gamma)$ for fixed values of other variables $\alpha, \beta+\gamma, \delta, \zeta$. 
Each line shows the result for different ($\alpha, \beta+\gamma, \delta, \zeta$). 
The magnetization is a convex upward function, and decreases with increasing the cerium concentration. 
This indicates that there is an optimum cerium concentration to effectively utilize abundant Ce element. 

\section{Conclusion} 
\label{sec:conclusion}

We have proposed a data-assimilation method in which a small number of accurate data and a large number of less accurate data are integrated. 
The method enables us to predict the behavior in the region where the former data are not available. 
Based on this method, we have developed a practical scheme to estimate the finite-temperature magnetization at arbitrary composition. 
We applied the scheme to {\nplcfcnb} and obtained the magnetization in the five-dimensional composition space. 
It is found that Co addition enhances the magnetization above 320 K.

\appendix
\section{Experimental procedure}
\label{sec:expdetails}
119 kinds of (Nd,Ce,La,Pr)$_{13.55}$-(Fe,Co,Ni)$_{80.54}$-B$_{5.91}$ (at\%) alloys were prepared by arc melting. These alloys were annealed at 1373 K for 24 h in Ar atmosphere. These alloys were pulverized and sorted into particles with diameters of $<$20 $\mu$m in an inert atmosphere to make magnetically anisotropic powder.
The powder density was determined using a pycnometer (Ulrtapyc1200e, Quantachrome Instruments, USA). The powder compositions were measured by ICP-AES (ICPS8100, Shimazu, Japan) and the main phase, 2-14-1 phase, ratio was calculated from the obtained composition. The magnetic physical properties were measured by using a vibrating sample magnetometer (PPMS EverCool II, QuantumDesign, USA) at a maximum applied field of 9 T.

The powder was mixed with an epoxy resin in a Cu container and solidified in a magnetic field of 1T to measure with VSM. The MH curve of magnetically easy and hard direction was measured in the temperature range of 300 K to 453 K. The anisotropy field ($H_A$) was detected by singular point detection (SPD) method~\cite{CABASSI2020107830} from MH curve of hard direction. 
When the anisotropy field is less than 9 T, less than the applied magnetic field of VSM, $H_A$ can be detected by SPD. If the anisotropy field is over 9 T, the MH curves for both of easy and hard direction were extrapolated in high magnetic field to obtain the intersection point, and the magnetic field at the intersection was made into $H_A$. On the other hand, the saturation magnetization ($J_s$) was estimated by the law of approach to saturation (LAS)~\cite{Akulov1931,hadjipanayis1981rare,kronmuller2003micromagnetism} from the MH curve of easy direction. 
When $H_A$ was higher than 9 T, which is the maximum applied field, the equation~(\ref{eq:exp1}) was used, and when it was sufficiently lower than 9 T, the equation~(\ref{eq:exp2}), (\ref{eq:exp3}) was used. 
$J_s$ is saturation magnetization and $b$ and $\chi_0$ are constants. 
The $\chi_0 H$ term is often referred to as the so-called paramagnetism-like term.~\cite{zhang2010law} 
Equation~(\ref{eq:exp2}), (\ref{eq:exp3}) dealing with the $\chi_0$ term is more accurate for measuring saturation magnetization, but there is a condition that the applied magnetic field is sufficiently larger than the anisotropic magnetic field. 
The saturation magnetization estimated by the LAS was divided by the main phase ratio of powder to obtain the saturation magnetization of the 2-14-1 phase in these composition, where the grain boundary phases other than the main phase were treated as paramagnetic phases.
\begin{equation}
J = J_s \left(1 - \frac{b}{H^2} \right) \;,
\label{eq:exp1}
\end{equation}

\begin{equation}
J = J_s \left(1 - \frac{b}{H^2} \right) + \chi_0 H \;,
\label{eq:exp2}
\end{equation}
\begin{equation}
\frac{dJ}{dH} = J_s \left( \frac{2b}{H^3} \right) + \chi_0 \;,
\label{eq:exp3}
\end{equation}

\section{First-principles calculation} 
\label{sec:calcdetails}
We perform density functional theory calculation 
following Korringa-Kohn-Rostoker (KKR)~\cite{Ko1947,KoRo1954} Green's function approach 
in the atomic sphere approximation (ASA) 
incorporating coherent potential approximation (CPA)~\cite{Sh1971,Ak1977}
as implemented in AkaiKKR.~\cite{Ak2010}
Continuous interpolation over $0\le \alpha, \beta, \gamma, \delta, \zeta \le 1$ for (Nd$_{1-\alpha-\beta-\gamma}$Pr$_{\alpha}$La$_{\beta}$Ce$_{\gamma}$)$_2$(Fe$_{1-\delta-\zeta}$Co$_\delta$Ni$_\zeta$)$_{14}$B
is done on the basis of KKR-CPA. 
The local spin density approximation
as parametrized by Moruzzi, Janak and Williams~\cite{MoJaWi1978} 
is adopted for the exchange-correlation energy functional. 
We set $l_{\rm max}=2$ putting all $4f$ electrons of the valence state
on the basis of open-core approximation. 
We assume each configuration of the electrons as Nd$^{3+}$, Pr$^{3+}$, and Ce$^{4+}$, respectively.
Contributions from $4f$ electrons of Nd and Pr to the magnetic moments
are restored manually. 

Intersite magnetic exchange couplings are calculated  
using the method developed by Liechtenstein {\it et al}.~\cite{LiKaAnGu1987} 
The Curie temperature is evaluated by solving the derived Heisenberg model in the mean-field approximation.

\section{Lattice parameters}
\label{sec:lattice}
We collected experimental lattice parameters of $R_2T_{14}$B ($R$=Nd, La, Ce, Pr;  $T$=Fe, Co, Ni) 
from literature~\cite{He1991} (Table~\ref{tab:lattice}(a)). 
However, some of them are not available. 
In order to evaluate the lattice parameters for these missing points, 
we performed first-principles calculation based on density functional theory~\cite{HoKo1964,KoSh1965}
in the generalized gradient approximation with 
Perdew-Burke-Ernzerhof formula~\cite{PeBuEr1996}
using VASP code~\cite{KrJo1999} (Table~\ref{tab:lattice}(b)).  
We integrated the calculated lattice parameters $a$ and $c$ with the above experimental data 
assuming the following simple form:
\begin{eqnarray}
a^{\rm comp} &=& \sum_{I} a_{I}^{\rm comp} \alpha^{i_\alpha} \beta^{i_\beta} \gamma^{i_\gamma} \delta^{i_\delta} \zeta^{i_\zeta}, \\
c^{\rm comp} &=& \sum_{I} a_{I}^{\rm comp} \alpha^{i_\alpha} \beta^{i_\beta} \gamma^{i_\gamma} \delta^{i_\delta} \zeta^{i_\zeta}, \\
a^{\rm expt} &=& C_a a^{\rm comp} + a_0 \;, \\
c^{\rm expt} &=& C_c c^{\rm comp} + c_0 \;,
\end{eqnarray}
where $I := (i_\alpha,i_\beta,i_\gamma,i_\delta,i_\zeta)$ runs under the condition $0 \leq \sum_j i_j \leq 1$.
Using the data-assimilation method explained in Sec.~\ref{sec:method}, 
we determined the coefficients $\{a_{I}^{\rm comp}\}, \{c_{I}^{\rm comp}\}, C_a, a_0, C_c, c_0$, 
from which $a^{\rm expt}$ and $c^{\rm expt}$ for the missing points are estimated  
(Table~\ref{tab:lattice}(c)). 
We interpolated lattice parameters for nonstoichiometric systems according to Vegard's law 
by using the obtained lattice parameters for stoichiometric systems.

\begin{table}
  \begin{tabular}{p{40pt}p{60pt}p{60pt}p{60pt}}
    \hline \hline
    (a) & Fe & Co & Ni \\
    \hline
    Nd & (8.80, 12.20) & (8.64, 11.86) & N/A \\
    La & (8.82, 12.34) & (8.67, 12.01) & N/A \\
    Ce & (8.76, 12.11) & N/A & N/A \\
    Pr & (8.80, 12.23) & (8.63, 11.86) & N/A \\
    \hline \hline  \\  
  \end{tabular}

  \begin{tabular}{p{40pt}p{60pt}p{60pt}p{60pt}}
    \hline \hline
    (b) & Fe & Co & Ni \\
    \hline
    Nd & (8.73, 12.07) & (8.58, 11.74) & (8.57, 11.75) \\
    La & (8.75, 12.16) & (8.61, 11.80) & (8.59, 11.79) \\
    Ce & (8.70, 11.96) & (8.55, 11.63) & (8.54, 11.62) \\
    Pr & (8.75, 12.11) & (8.59, 11.77) & (8.59, 11.77) \\
    \hline \hline  \\  
  \end{tabular}

  \begin{tabular}{p{40pt}p{60pt}p{60pt}p{60pt}}
    \hline \hline
    (c) & Fe & Co & Ni \\
    \hline
    Nd &  &  & (8.62, 11.89) \\
    La &  &  & (8.65, 11.93) \\
    Ce & & (8.61, 11.77)& (8.60, 11.75)  \\
    Pr &  &  & (8.64, 11.91)) \\
    \hline \hline  \\  
  \end{tabular}

\caption{Lattice parameters $(a, c)$ of $R_2T_{14}$B for $R$=Nd, La, Ce, Pr and $T$=Fe, Co, Ni. 
(a) Experimental values taken from Ref.~\onlinecite{He1991}, 
(b) computational results, and 
(c) values obtained by the data-assimilation technique to complement the unavailable values of (a).
}
\label{tab:lattice}
\end{table}

\begin{acknowledgments}
This work was supported by 
MEXT as “Program for Promoting Researches on the Supercomputer Fugaku” (DPMSD). 
Part of the computation in this work was performed using the facilities of the Supercomputer Center of the Institute for Solid State Physics at the University of Tokyo, and computational resources of the HPCI system through the HPCI System Research Project (Project ID:hp200125).
\end{acknowledgments}

\bibliography{./Reference.bib}

\end{document}